\documentclass[pre,twocolumn,superscriptaddress,showpacs,floatfix,aps]{revtex4-2} 
\usepackage{hyperref}
\usepackage{dcolumn} 
\usepackage{bm} 
\usepackage{amssymb} 
\usepackage{amsmath}
\usepackage{color} 
\usepackage{graphics}
\usepackage[]{natbib}

\usepackage{epsfig}
\usepackage[version=4]{mhchem}
\begin{document}

\title{Thermodynamics, structure and dynamics of cylindrically confined hard spheres: The role of excess helical twist.}

\author{Mahdi Zarif}
\email{m_zarif@sbu.ac.ir}
\affiliation{Department of Physical and Computational Chemistry, Shahid Beheshti University, Tehran 19839-9411, Iran.}

\author{Richard K. Bowles}
\email{richard.bowles@usask.ca}
\affiliation{Department of Chemistry, University of Saskatchewan, Saskatoon, SK,S7N 5C9 , Canada.}
\affiliation{Centre for Quantum Topology and its Applications (quanTA), University of Saskatchewan, SK S7N 5E6, Canada.}

\date{\today}

\begin{abstract}
Hard spheres confined to narrow quasi-one-dimensional cylindrical channels form perfect helical structures at close packing. Here, we use molecular dynamics simulation to show that the thermodynamics, structure and dynamics of the fluid below close packing are dominated by the presence of topological defects that reverse the local twist direction of the helix. When compressed from a random, low density state, or decompressed from high density ordered states with zero excess helical twist, the system equilibrates to an achiral fluid that exhibits two heat capacity maxima along the equation of state. The low density heat capacity maximum corresponds to the onset of helix formation and the high density maximum occurs when the system rapidly loses defects in a Schottky-like anomaly. The local twist auto-correlation function in the achiral fluid exhibits a stretched exponential decay and the structural relaxation times undergo a fragile-to-strong crossover located at the high density heat capacity maximum. We also study the effect of excess helical twist by using initial starting configurations consisting of two helical domains with opposite twist directions of different lengths. This leads to the formation of topologically protected states that are characterized by the presence of loosely bound defect pairs which become more tightly bound with increasing excess helical twist. The local twist auto-correlation function in the chiral fluid decays as a power law at long times. The possible kinetic or thermodynamic origin of this topological protection is discussed.
\end{abstract}

\maketitle
 
\section{Introduction}
\label{sec:intro}
Topological order and topologically protected states usually appear in correlated quantum systems~\cite{Senthil_Symmetry_1014} but there is growing evidence to suggest that they can also play an important role in the properties of classical systems~\cite{Kane_Topological_2013,Loehr_topological_2016,Pedro_Topological_2019}. For example, Zygmunt et al.~\cite{Zygmunt_Topological_2019} found that dense packings of anisotropic colloids form topological phases that retain near perfect order below close packing, demonstrating a form of classical topological protection. The origin of  topological order in these systems differs from those of their quantum counterparts and arises from the organization of particle contacts within the unit cell of the packing which suggests other colloidal systems may exhibit similar phenomena. 

The geometric confinement of hard sphere particles to narrow quasi-one-dimensional (q1d) channels leads to the spontaneous formation of helical structures~\cite{Erickson_tubular_1973,Harris_Tubular_1980,Pickett_Spontaneous_2000} ranging from simple single helices through to multi-stranded helices with slip or staggered structures, depending on the channel diameter. When the channel diameter is sufficiently narrow, all the particles in the dense packings contact the channel wall, which allows their structures, and the transitions between them, to be described in terms of phyllotactic disk arrangements on a plane~\cite{Mughal_Phyllotatic_2011,Mughal_Dense_2012,Mughal_Theory_2014,Fu_hard_2016}. A variety of new structures arise as the channel becomes wide enough for particles to enter the core of the packings~\cite{Duran_Ordering_2009,Fu_Assembly_2017}, and eventually, the appearance of  bulk face-centred-cubic (FCC) crystal packing arrangements leads to the formation of complex core-shell structures~\cite{Zhu_Entropy_2021}. A similar array of structures have been found in confined quasi-1d soft sphere systems~\cite{Oguz_Helicity_2011,Winkelmann_Simulation_2017,Mughal_Columnar_2018}, and have been observed experimentally in molecular nanotube systems~\cite{Michelson_Packing_2003,Khlobystov_Observation_2004,Yao_Stable_2016}, colloidal particles~\cite{Lohr_Helical_2010,jiang_Helical_2013,Fu_Assembly_2017,Jimenez_Self_2021} and macroscopic, athermal systems~\cite{Meagher_Experimental_2015}. Introducing particle shape anisotropy then broadens the range of structural motifs formed under cylindrical confinement and can generate new chirality elements~\cite{Jin_Shape_2020,Wan_Shape_2022}.

Colloidal crystals have applications in photonics~\cite{Vlasov_On_Chip_2001,Zhao_Spherical_2014,Lei_Self_2018,Wan_Randomness_2020} and jammed helical packings of hard spheres have been shown to exhibit chiral photonic properties~\cite{Liu_Chiral_2023}. However, it is not known if these structures remain chiral below close packing. Quasi-one-dimensional systems with short ranged interactions generally do not exhibit phase transitions~\cite{vanhove_Sur_1950,Ruelle_Statistical_1968} because there is always an entropic advantage to introducing a defect into the system that overcomes the energetic cost of the defect in the thermodynamic limit~\cite{landau}.  Randomly distributed defects would break up the helical structure, leading to an achiral fluid. Nevertheless, there are circumstances where phase transitions can occur in quasi-one dimensional systems~\cite{Cuesta_General_2004,Saryal_Multiple_2018} and phase transitions can arise in 1d systems when the particle interactions become long-range and the defects have a topological character~\cite{Dyson_Existence_1969,Thouless_Long_1969,Kosterlitz_Nobel_2017}. 


The current work examines the role helical topology plays in the structural, thermodynamic and dynamic properties of a system of hard spheres confined to a narrow, q1d channel, focusing on a system that has the simplest, perfect single helix ground state~\cite{Yamchi_Helical_2015,Chan_Densest_2019,Zarif_Inherent_2021}. To capture the topological properties of the fluid we calculate the local twist direction around each particle (see Methods) allowing us to identify sections of helix and structural ``defects" in fluid where the twist direction of the helix changes, i.e. between left ($\mathsf{M}$) and right ($\mathsf{P}$) twist directions. The method provides an approximate mapping of a fluid configuration to its local jammed, or inherent structure~\cite{Stillinger_Systematic_1964,Stillinger_Hidden_1982,Ashwin_Inherent_2013}, which helps us relate the thermodynamics and dynamics of the fluid to its underlying topological structure through the number and distribution of the helical defects. We find that an equilibrium achiral fluid readily forms when compressed from a random, low density state, or decompressed from high density ordered states with zero excess helical twist. The properties of the achiral fluid are consistent with those expected from a q1d hard sphere system~\cite{Yamchi_Inherent_2015,Fu_Assembly_2017}. However, introducing an excess of helical twist into a high density initial configuration leads to the formation of structurally distinct, topologically protected fluid states at intermediate densities, with properties that depend on the degree of excess and are characterized by the appearance of loosely bound defect pairs. At low densities, these states decay into the achiral fluid. The remainder of the paper is organized as follows: Section~\ref{sec:model} describes the model and simulations methods and Section~\ref{sec:res} describes our results for the systems with zero and non-zero excess helical twist. Our discussion and conclusions are contained in Sections~\ref{sec:disc} and ~\ref{sec:conc}, respectively.

\section{Methods}
\label{sec:model}
\subsection{Model}
We study a system of $N$ hard spheres with diameter $\sigma$, confined in a cylindrical narrow channel of length \textit{L} with channel diameter
$H_{d}/\sigma=1.95$,  which ensures spheres can only contact their first and second neighbours in either direction along the channel. The particle--particle and particle--wall interaction potentials are given by,
\begin{equation}
U(r_{ij})= \left\{\begin{matrix}
0 \;\;\; & r_{ij} \geq \sigma \;\;\;\;\;\;\;\;\\
\infty \;\;\; & r_{ij} <  \sigma \;\;\;\;\;\;\;\;
\end{matrix}\right.,
\label{potential}
\end{equation}
\begin {equation}
U_{w}(r_{i})= \left\{\begin{matrix}
0 \;\;\; & |r_{xy}| \leq  \left | H_{0}/2 \right | \\
\infty \;\;\; & \textup{otherwise}
\end{matrix}\right.,
\label{wpotential}
\end{equation}           
respectively, where $r_{ij}=\left | \mathbf{r_{i}-r_{j}} \right |$ is the distance between particles, $|r_{xy}|$ is the magnitude of position vector for a particle perpendicular to the wall where the centre of the cylinder is located at $x = y = 0$ and the longitudinal direction of the channel extends in the $z$ direction. The volume accessible to the particles' centres is $A_{0}L=\pi L (H_{0}/2)^{2}$, where $A_0$ is the cross sectional area of the channel accessible to the particle centres and  $H_{0}=H_{d} - \sigma$.  The occupied volume fraction $\phi=2 N \sigma ^{3}/\left (3LH^{2}_{d}\right)$. 

\subsection{Molecular Dynamics Simulation}
Our study uses event driven molecular dynamics (MD), performed in the canonical ensemble ($N,V,T$), with systems containing $N=10^{4}$ particles and $H_d/\sigma=1.95$.  The unit of time for the simulation is given by $\sigma \sqrt {m/k_BT}$, where $k_{\text{B}}$ is Boltzmann's constant and $m$ is the mass of a particle, which is set to unity. Particles are assigned random velocities at the beginning of the run, scaled to ensure $k_BT=1$, and velocity rescaling is used to maintain the temperature. 

Establishing equilibrium in this system is difficult at high densities. To help identify the equilibrium branch of the fluid and to explore the properties of the non-equilibrium region of the phase diagram, we use a number of different starting conditions and protocols. The {\it Compression } protocol begins with the particles placed in a linear lattice at $\phi=0.01$. A MD simulation is performed at the initial $\phi$, before the system is compressed to a higher occupied volume fraction using a modified version of the Lubachevsky and Stillinger event-driven algorithm~\cite{Lub_Geometric_1990} that expands the particles and channel diameter such that $H_{d}/\sigma$ remains constant.  A compression rate of $d\sigma/dt=0.001$ is used. Depending on $\phi$, simulation runs of between $2\times 10^5N$ and $6\times 10^5N$ collisions are performed at each state point, with equilibrium data being collected over the last $1\times 10^5N$ collisions of the trajectory. Translational periodic boundary conditions are used, but twist boundary conditions are not employed.

We also decompress the system from high densities using a number of different starting conditions. We generate a series of {\it Defect Crystals} that consist of identical helical sections, containing $n_h$ particles, separated by defects that change the twist direction. The properties of the helix in each section, such as the angles and distances between particles, are obtained by minimizing $L/N$ in a model for helical hard sphere packings with defects~\cite{Zarif_Inherent_2021}. The length of the channel and the longitudinal particle positions were then scaled to generate a starting configuration with $\phi=0.395$.  We study systems with $n_h=50,100,500,1000,2500$, which contain defect fractions $\theta=n_{d}/N= 0.02, 0.01, 0.002, 0.001,$ and $0.0004$, respectively, where $n_{d}$ is the number of defects in the system. The number of sections with right ($\mathsf{P}$) and left ($\mathsf{M}$) hand twist directions is the same so the twist excess $\chi=0$ (see Section~\ref{sec:hstruc} for details). 

Finally, we create a series of starting configurations, each with a single {\it Pair} of defects ($\theta=0.0002$) arranged to form two helical sections of different lengths, giving rise to an initial twist excess, $\chi_I$. We study systems with $\chi_I=0.0, 0.2, 0.4, 0.6, 0.8$ and $0.9996$, where this last state point consists of a system with one helical section containing  9998 particles and the other containing two particles. Note, we cannot have a system with $\chi_I=1.0$ because of the inherent twist associated with the perfect helical structure and the fact that we only use translational periodic boundaries. The system is decompressed between $\phi$ by increasing $L$ and rescaling the particle positions along $z$, while maintaining $\sigma$ and $H_d$ fixed. Twenty independent simulations, obtained by assigning different velocity distributions, are performed for each starting condition and results are averaged over all simulations.

\subsection{Heat Capacity}
We calculate a number of thermodynamic and structural properties of the fluid. The constant pressure heat capacity for the system is given by,
 \begin{equation}
 \frac{C_p}{Nk}=\frac{1}{Nk}\left(\frac{\partial H}{\partial T}\right)_{P_L}=\frac{3}{2}+\frac{Z}{1+\left(\frac{\partial \ln Z}{\partial \ln \phi}\right)_{T,H_d}}\mbox{,}\\
 \label{eq:cp}
 \end{equation}
where $H$ is the enthalpy, $Z=P_LA_0L/NkT$ is the compressibility factor.

\subsection{Helical Structure}
\label{sec:hstruc}
To examine the helical structure of the fluid, we identify a local helical twist direction for each atom $i$ based on the signed volume of the tetrahedron given by;
\begin{equation}
v_{tet}(i)=\frac{\mathbf{a}\cdot\mathbf{b}\times\mathbf{c}}{6}\mbox{,}\\
\label{seq:vtet}
\end{equation}
where $\mathbf{a}$, $\mathbf{b}$ and $\mathbf{c}$ are the position vectors for particles $i-1$, $i$ and $i+1$, relative to particle $i+2$, respectively. Successive particles with the same sign for $v_{tet}(i)$ have the same helical twist direction, which allows us to identify the length of a helical section. Defects, located between helical sections with opposite twist directions, occur when $v_{tet}$ changes sign. The method has been used to study helical structure in the jammed packings of this system~\cite{Zarif_Inherent_2021}, where the small magnitude of $|v_{tet}|$ at a defect was also used. In the fluid, particles adopt configuration with a broad distribution of $|v_{tet}|$ so this is no longer possible. The current work identifies the location of the defects just using the sign changes of $v_{tet}(i)$. Our analysis of the helical structure in the fluid represents an approximate mapping of a fluid configuration to a nearby jammed structure, effectively providing an inherent structure landscape~\cite{Stillinger_Systematic_1964,Stillinger_Hidden_1982,Ashwin_Inherent_2013} description for the system. However, we find cases where a pair of defects are located on neighbouring particles, which represents an unstable environment that leads to defect annihilation when compressed to jamming~\cite{Yamchi_Helical_2015}. 

We also calculate the probability, $P(n)$, of finding a helical section containing $n$ particles and the excess helical twist, 
\begin{equation}
\chi=\frac{N_{\mathsf{P}}-N_{\mathsf{M}}}{N}\mbox{,}\\
\label{eq:excess}
\end{equation}
where $N_{\mathsf{P}}$ and $N_{\mathsf{M}}$ are the number of particles with right and left hand local twist directions, respectively. Assuming the number and distribution of the defects in the fluid provide an effective instantaneous map of a configuration to its inherent structure, we can follow the evolution of the system through the inherent structure landscape as a function of density. 

\subsection{Structural Relaxation}
Structural relaxation in the system occurs through the creation, diffusion and elimination of defects because these change the local direction of helical twist. To measure structural relaxation, we define a local twist auto-correlation function,
\begin{equation}
g_0=\left<\frac{v_{tet}(i,0)}{|v_{tet}(i,0)|}\frac{v_{tet}(i,t)}{|v_{tet}(i,t)|}\right>\mbox{,}\\
\label{eq:g0}
\end{equation}
where the average is taken over all particles and multiple time origins have been used.

\section{Results}
\label{sec:res}
\subsection{Zero Excess}
\label{sec:zero}
\begin{figure}[h]
\includegraphics[width=3in]{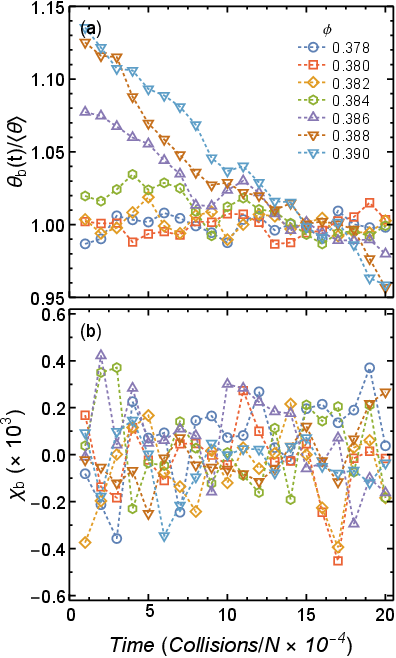}
\caption{\textbf{Compression Equilibrium.} (a) Relative block averaged defect fraction, $\theta_b(t)/\langle\theta\rangle$ and (b) block averaged excess twist, $\chi_b(t)$, as a function of time, for the compression simulations at $\phi=0.378-0.390$.}
\label{fig:compequil}
\end{figure}

The compression and {\it defect crystal} systems are considered to have zero excess because the initial condition for the simulations has no excess in the helical twist and the measured average excess remains zero for all densities studied. We begin by identifying the density regime where the zero excess simulations are in equilibrium over the time scale of our simulations. Considering the {\it Compression} simulations first: Figure~\ref{fig:compequil}(a) shows the block average values of the defect fraction, $\theta_b(t)$ as a function of time over the last $2\times 10^5$ collision$/N$ of the simulation, relative to the average defect fraction, $\langle\theta\rangle$, obtained over the last $10^5$ collisions$/N$. The block averages are calculated over $10^4$ collisions$/N$. If the system is in equilibrium, we expect $\theta_b(t)/\langle\theta\rangle$ to fluctuate around a value of one over the last half of the plot. Densities $\phi=0.378-0.382$ are clearly in equilibrium well before the region we calculate our averages, and simulations at $\phi=0.384$ reach equilibrium. Simulations at $\phi=0.386$ appear to be nearing equilibrium.  However, simulations at higher densities are clearly still evolving to lower defect fractions, even in the last section of the run, and are out of equilibrium. The block average excess fluctuates within $0.02\%$ of zero at all densities (Fig.~\ref{fig:compequil}(b)).

\begin{figure}[t]
\includegraphics[width=3in]{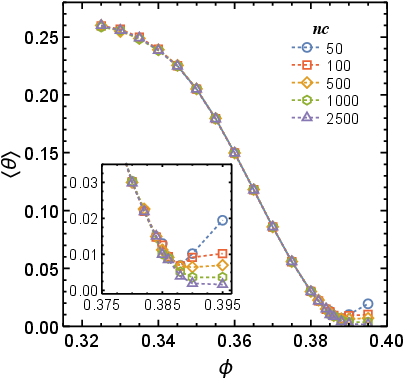}
\caption{\textbf{Defect Crystal Equilibrium.} Average defect fraction as function of $\phi$ for {\it Defect Crystals} with $n_h=50-2500$. Insert: Expanded high density region of the main plot.}
\label{fig:dcequil}
\end{figure}

The {\it Defect Crystal} simulations start at high density and are then decompressed. However, {\it Defect Crystals} with higher initial defect fractions (smaller $n_h$) have a lower jamming density, $\phi_J$. Hence, they have less free volume available for rearrangement and structural relaxation. Figure~\ref{fig:dcequil} shows that the {\it Defect Crystals} with small $n_h$ initially retain a high value of $\langle\theta\rangle$ at high densities, but all systems converge to the same values below $\phi=0.385$, suggesting the properties of the system become independent of their initial starting point.
Given the high level of consistency between different {\it Defect Crystal} simulations, below $\phi=0.385$ we calculate averages over all $n_h$ and present the results as a single data set in the remaining analysis. Furthermore, the trajectory data for these simulations are similar to those of the $Compression$ simulations, and all systems retain a zero excess at all densities.

\begin{figure}[t]
\includegraphics[width=3in]{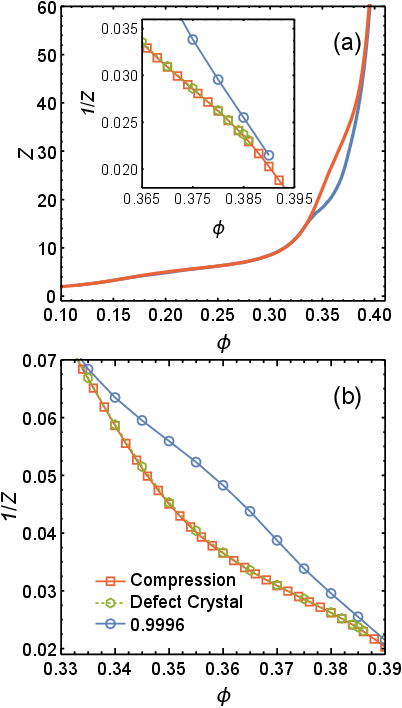}
\caption{\textbf{Equation of state.} (a) $Z$ for {\it Compression} and $\chi_I=0.9996$ decompression branches over full range of $\phi$. Insert: $1/Z$ in the high density region (b) $1/Z$  for {\it Compression}, {\it Defect Crystal} and  {\it Pairs} ($\chi_I=0.9996$) in the bifurcation region. }
\label{fig:eos}
\end{figure}

Figure~\ref{fig:eos}(a) shows the compressibility factor for the system over a wide range of $\phi$, for the compression branch and for the $\chi_I=0.9996$ decompression branch. Below $\phi\approx0.33$, all protocols follow the same equation of state (EOS), which exhibits a shoulder centred around $\phi\approx 0.25$ and approaches the gas behaviour in the low density limit. Above $\phi\approx0.33$ there is a bifurcation in the EOS's, where the {\it Pair} protocols no longer converge on the time scale of the simulations. These will be described in detail in Section~\ref{sec:eht}. Both the {\it Compression} and {\it Defect Crystals} protocols follow the same path at high density (Figure~\ref{fig:eos}(b)), although, we start to see a hint of separation between the two protocols at the highest densities as highlighted in the insert. The EOS of the zero excess ({\it Compression/Defect Crystal}) branch obtained by our MD simulations follows that obtained by Hu et. al.~\cite{Hu_Correlation_2018} using a transfer matrix analysis.

\begin{figure}[t]
\includegraphics[width=3in]{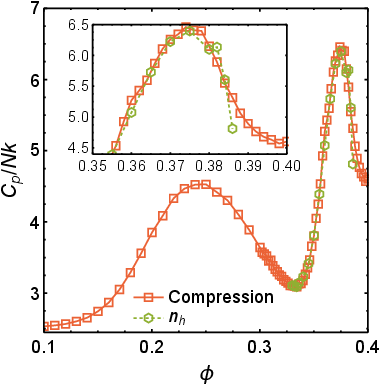}
\caption{\textbf{Zero Excess Heat capacity.} $C_p/Nk$ as a function of  $\phi$. Insert: $C_p$ maximum in the high density region.}
\label{fig:cp}
\end{figure}

\begin{figure}[t]
\includegraphics[width=3in]{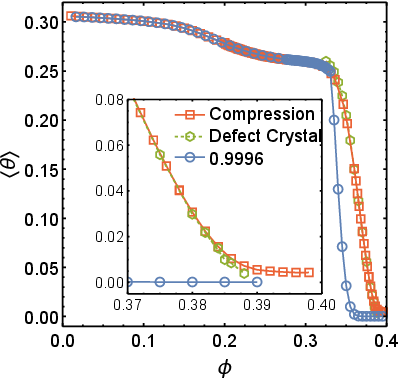}
\caption{\textbf{Defect fraction.} $\langle\theta\rangle$ as a function of $\phi$ for {\it Compression}, {\it Defect Crystal} and $\chi_I=0.9996$. Insert: High density region.}
\label{fig:defects}
\end{figure}

The heat capacity calculated along the zero excess branch exhibits two heat capacity maxima (Fig.~\ref{fig:cp}), suggesting the system goes through structural changes with increasing $\phi$, one at low densities with a peak centred at $\phi\approx 0.25$ and the second at higher densities, above the bifurcation point in the EOS, with a peak centred around $\phi\approx 0.375$. We characterize the structural evolution of the system by following $\langle\theta\rangle$ as a function of $\phi$ (Fig.~\ref{fig:defects}) and comparing our results with the inherent structures of the model~\cite{Zarif_Inherent_2021}. In the ideal gas limit, the system is expected to sample basins in the inherent structure landscape near the maximum of the distribution~\cite{Ashwin_Inherent_2013} where $\theta= (5-\sqrt{5})/10\approx0.276$~\cite{Yamchi_Helical_2015}. Our simulations find  $\langle\theta\rangle\approx 0.3$, which is only marginally higher. With a high fraction of defects, the number of spheres between the defects is small and the particles tend to adopt linear or zig-zag arrangements. As the density increases, the number of defects decreases rapidly before reaching a plateau at $\theta\approx 0.25$ where the average defect separation is four particles. This corresponds to the smallest section capable of forming part of a helical twist and coincides with the low density heat capacity maximum. The onset of helical structure in this system was previously observed in simulation~\cite{Fu_Assembly_2017,Hu_Correlation_2018}.

The structure of the fluid along the rest of the compression branch consists of loosely organized sections of helix, separated by defects, where all the particles in a given helical section have the same local twist direction and the twist alternates between $\mathsf{M}$  and $\mathsf{P}$ directions. With increasing $\phi$, $\langle\theta\rangle$ begins to rapidly decrease again as the fluid continues to move to basins on the inherent structure landscape associated with jammed structures with higher $\phi_J$, maximizing its total entropy by trading configurational entropy for increased vibrational entropy.  Eventually the system finds its way toward the bottom of the landscape where there are very few basins, characterized by a small number of defects, and $\langle\theta\rangle$ begins to plateau. This ordering process  coincides with the high density $C_p$ maximum. At the highest densities, we see the {\it Compression} data separate from the {\it Defect Crystals} as the system falls out of equilibrium on the time scale of our simulations.

\begin{figure}[t]
\includegraphics[width=3in]{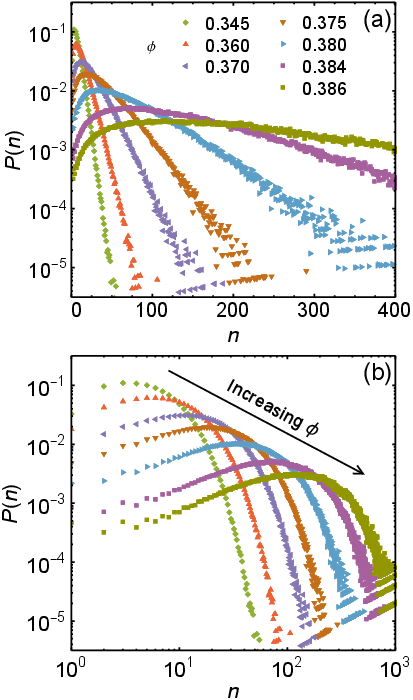}
\caption{\textbf{Helix size distribution.} (a) Log-Linear plot of $P(n)$ as a function of  helix size, $n$, for different $\phi$. (b) Log-Log plot of the same data.}
\label{fig:hd}
\end{figure}

The topological nature of the fluid structure means that defects are eliminated in pairs leading to the formation of larger helical sections. Furthermore, as there is no preference for one helical twist direction over another, the defects eliminate randomly and $\langle \chi \rangle$ is zero over the entire {\it Compression} and {\it Defect Crystal} branch of the EOS. Figure~\ref{fig:hd}(a) shows that $P(n)$, the probability distribution for helical section sizes, decays exponentially with helical section size at intermediate pressures, before the $C_p$ maximum. This is consistent with the general form predicted for a random distribution of helical  sections, with the exception of the appearance of the single particle section, which is not accounted for in the random model. As the high density $C_p$ maximum is approached, the form of the distribution changes. Figure~\ref{fig:hd}(b) shows that the small helix distribution follows a power law behaviour, which grows as the maximum in $P(n)$ moves to larger $n$ with increasing $\phi$. We also see an oscillation in the probabilities for small odd and even sized helical sections, which was also observed in the inherent structures of the model~\cite{Zarif_Inherent_2021}. Notably, the large $n$ tails retain their exponential character.

\begin{figure}[t]
\includegraphics[width=3in]{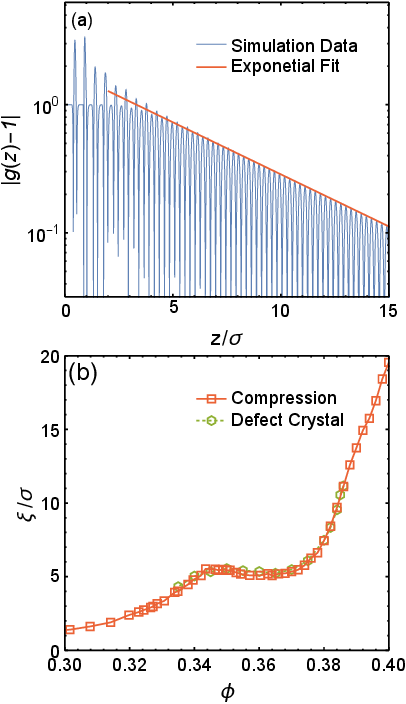}
\caption{\textbf{Translational Correlation.} (a) Log plot of the longitudinal pair correlation function $|g(z)-1|$ as a function of distance along the channel, $z\sigma$ obtained at  $\phi=0.378$ from the {\it Compression} branch (simulation data - blue line) and the exponential fit to the data peaks (red line). (b) Translational correlation length, $\xi/\sigma$, as a function of $\phi$, for {\it Compression} and {\it Defect Crystal} protocols.}
\label{fig:tc}
\end{figure}

Figure~\ref{fig:tc}(a) shows that the longitudinal pair correlation function decays exponentially, beyond the small $n$ region, as expected for a q1d fluid with no long range translational order, which also allows us to extract a translational correlation length, $\xi/\sigma$, by fitting the peaks. Figure~\ref{fig:tc}(b) shows $\xi/\sigma$ grows slowly at low $\phi$ before reaching a plateau, or even a weak maximum, just above bifurcation density. The translational correlation function then begins to grow again above $\phi\approx 0.36$, but the growth appears to slow at the higher densities, where the system begins to fall out of equilibrium.

\begin{figure}[t]
\includegraphics[width=3in]{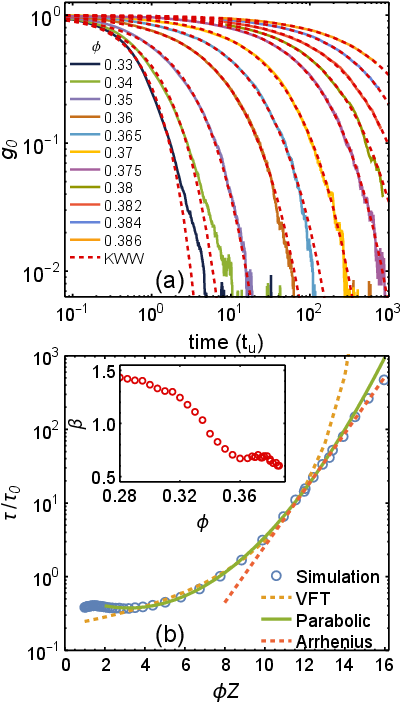}
\caption{\textbf{Orientational Correlation.} (a) Log-log plot of the orientational correlation function, $g_0$, as a function of time for different $\phi$ (solid lines) and the fit to KWW function (dashed lines). (b) Orientational Relaxation of time, $\tau/\tau_0$, as a function of $\phi Z$. Insert: Stretched exponential exponent $\beta$ as a function of $\phi$.}
\label{fig:oc}
\end{figure}

Figure~\ref{fig:oc}(a) shows that along the compression branch of the EOS, the local twist correlation function, $g_0$, decays to zero on the time scale of our simulations over a wide range of $\phi$, including those densities at and above the high density $C_p$ maximum, highlighting the fluid-like nature of the system. The time dependence of $g_0$ also fits the Kohlraush-Williams-Watts (KWW) function~\cite{Kohlrausch_Theories_1854,Williams_Non_1970}, $\exp\left[-(t/t_r)^{\beta}\right]$,  where $t$ is time, $t_r$ is a time constant and $\beta$ is an exponent that characterizes the nature of the dynamics in a variety of supercooled liquids~\cite{Debenedetti_Supercooled_2003,Wu_Heterogeneous_2016}. At low $\phi$, we find $\beta > 1$ (see insert Fig.~\ref{fig:oc}(b)), which indicates the relaxation follows a compressed exponential decay, suggesting the fluid relaxes through a combination of ballistic and diffusive dynamics~\cite{Bouchaud_comp_2008}. When the particles are well separated along the channel, they tend to collide with the channel wall, reversing direction, before particle-particle collisions occur. This can result in a reversal of the sign of $v_{tet}$ as the particles cross the channel relative to the other particles that form the tetrahedron, and $g_0$ exhibits negative correlations after a short time (not shown), similar to those observed in the velocity auto-correlation function of bulk hard spheres. With increasing $\phi$, particle-particle collisions tend to cage a given particle, maintaining the sign of $v_{tet}$, leading to diffusive behaviour and we see $\beta$  decreases linearly. Interestingly, we see a change of slope in $\beta$ at $\phi\approx 0.33$, which coincides with density where the helical structure of the liquid begins to develop. At high $\phi$, $\beta$ plateaus at $\approx 0.65$, where values of $\beta < 1$  indicate a stretched exponential relaxation that is characteristic of slow glassy dynamics.

Supercooled liquids exhibit a range of behaviour for the temperature dependence of their relaxation times, $\tau$~\cite{Angell_Formation_1995}. In a strong liquid, $\tau$ has an Arrhenius temperature dependence, $\tau\sim\exp(A/T)$ where $A$ is constant. Fragile liquids exhibit super Arrhenius behaviour that can be described by a number of models, such as the Vogel-Fulcher-Tammann  (VFT) equation~\cite{Vogel_Das_1921,Fulcher_Analysis_1925,Tammann_Die_1926}, $\tau\sim\exp\left[A/(T-T_0)\right]$ which predicts a divergence at a finite temperature $T_0$, and the parabolic law~\cite{Elmatad_Corresponding_2009,Elmatad_Corresponding_11_2010}, $\tau\sim\exp[A/T^2]$, which predicts no divergence and arises from a facilitated dynamics description of glassy fluids. A two-dimensional (2d) system of hard discs confined to a narrow channel~\cite{Yamchi_Fragile_2012} even exhibits a crossover from fragile to strong fluid behaviour, located at the $C_p$ maximum, where the fragile behaviour at low $\phi$ occurs because defect-defect annihilation creates irreversible particle rearrangements that form more stable states. As $\phi$ increases, the defects become rare and structural relaxation proceeds through the simple hopping events of isolated defects, leading to strong fluid behaviour.  

To examine the nature of the dynamics in the current system, $\tau$ is defined as the time required for $g_0$ to decay to 0.2.  For hard spheres, $\phi PV$ is a constant along an isobar so $\phi Z$ varies as $\sim1/T$, and a log plot of $\tau$ vs  $\phi Z$  represents an effective Arrhenius plot for the relaxation times (see Fig.~\ref{fig:oc}(b)). The VFT and parabolic equations are fit to the data obtained  from the fragile fluid  densities, between the low and high pressure $C_p$ maxima on the compression EOS. The Arrhenius equation is fit over the data from $\phi$ at and above the high pressure $C_p$ maximum, up to $\phi=0.386$ where the system is in equilibrium. The VFT and Arrhenius equations fit the data over the fragile and strong regions respectively, and extending the data fitting region for either equation leads to a decrease  in the quality of the fits, suggesting this system may well exhibit a fragile-strong dynamic crossover, similar to that observed in the 2d confined fluid. The parabolic law, which does not contain a fragile-strong crossover, describes the relaxation times over most of the range studied and only deviates at the highest densities, despite only being fit using the data from below the $C_p$ maximum.

\subsection{Excess Helical Twist}
\label{sec:eht}
\begin{figure}[t]
\includegraphics[width=3in]{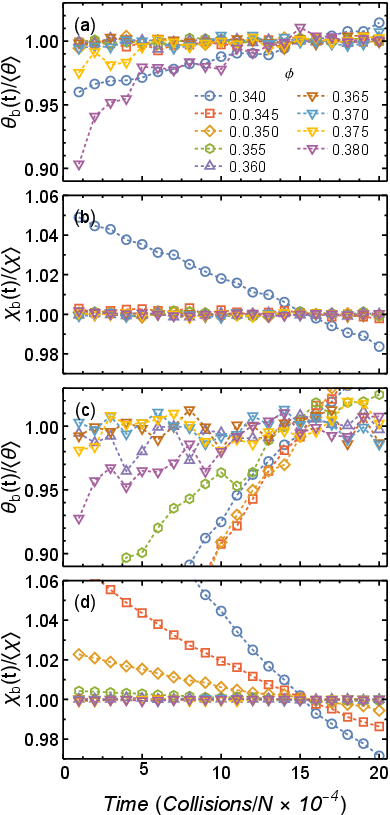}
\caption{\textbf{Excess Equilibrium.} Relative block averages for (a) $\theta_b$ for $\chi_I=0.4$, (b) $\chi_b$ for  $\chi_I=0.4$ (c) $\theta_b$ for $\chi_I=0.8$, (d) $\chi_b$ for  $\chi_I=0.8$, as a function of $\phi$.}
\label{fig:excesstraj}
\end{figure}

The {\it Pairs} protocol imposes a non-uniform distribution of the twist excess at high density by introducing a pair of defects into the initial configuration so the system has two helical sections with opposite twist directions. Positioning the defect at different separations gives rise to the values of $\chi_I$. The system is then decompressed step wise as described in the Methods section. Figure~\ref{fig:excesstraj} shows the block averaged defect fraction and excess helical twist, relative to their average values over the final $10^5$ collisions per particle of the trajectory at different densities. At densities higher than $\phi=0.380$ (not shown here) $\theta_b$ increases over the entire trajectory, showing the system is out of equilibrium. At lower $\phi$, we see  $\theta_b$ increases with time during the early part of the trajectory, but plateaus at later times suggesting the system is nearing some form of equilibrium.  Eventually we reach a regime where $\theta_b/\langle\theta\rangle$ fluctuates around unity even before we reach the region of the trajectory used for our analysis. However, $\theta_b$ then becomes time dependent again at lower $\phi$, over the entire trajectory, indicating the system has fallen out of equilibrium again. The excess helical twist appears to be time independent at high densities, but also becomes time dependent at the same low $\phi$ where the defect concentration falls out of equilibrium. In comparing the results for different $\chi_I$, we note the $\chi_I=0.8$ system (Fig.~\ref{fig:excesstraj} (c-d)) falls out of equilibrium at $\phi\approx0.355$, but the $\chi_I=0.4$  (Fig.~\ref{fig:excesstraj} (a-b)) remains stable until $\phi=0.340$. This is a general trend with the systems with lower $\chi_I$ remaining stable to lower $\phi$.

\begin{figure}[t]
\includegraphics[width=3in]{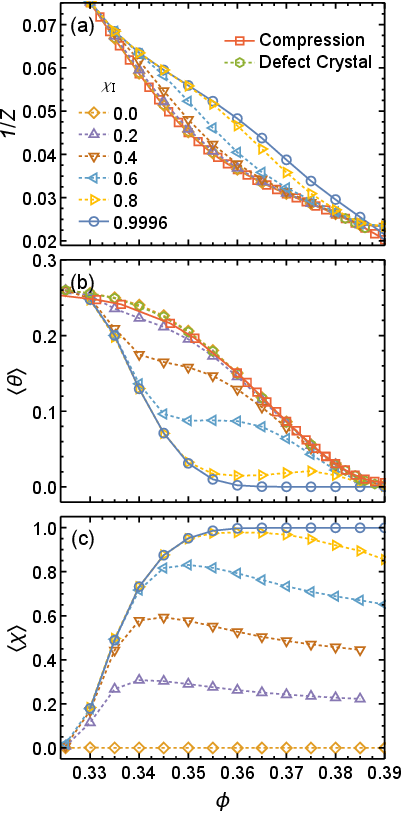}
\caption{\textbf{Excess Properties} (a) $1/Z$, (b) $\langle\theta\rangle$ and (c) $\langle\chi\rangle$ with different $\chi_I$ as a function of $\phi$.}
\label{fig:propexcess}
\end{figure}

Figure~\ref{fig:propexcess} summarizes the EOS, defect fraction and measured helical excess of the systems studied using the $\it Pair$ protocol as a function of $\phi$, for different values of the initial excess. It is interesting to note that at high densities, the EOS of the {\it Pair} simulations closely follow those of the {\it Compression} and {\it Defect Crystal} protocols, before branching off in an almost linear fashion at intermediate densities, with the lower $\chi_I$ systems deviating to a lesser degree (Fig.\ref{fig:propexcess}(a)). The EOS converge again at $\phi\approx 0.33$. Here it is important to re-emphasize that the {\it Pair} simulations fall out of equilibrium at some $\phi$ that decreases with decreasing $\chi_I$. The $\chi_I=0.9996$ case falls out of equilibrium below $\phi\approx0.360$ and $1/Z$ begins to move lower. Similarly, $\chi_I=0.8$ falls out of equilibrium at $\phi\approx 0.355$ and its $1/Z$ begins to move lower, and so on.

The density dependence of the defect fraction (Fig.\ref{fig:propexcess}(b)) exhibits similar characteristics to those observed in the EOS. At high densities, $\langle\theta\rangle$ for the {\it Pair} protocol again follows the  {\it Compression} and {\it Defect Crystal} protocols before deviating at intermediate $\phi$. In particular, we see $\langle\theta\rangle$ plateau or go through a maximum and begin to decrease for $\chi_I=0.6$ and $0.8$ respectively. The defect fraction begins to increase again at lower densities. However, $\langle\chi\rangle$ behaves differently, increasing from its initial high density value as $\phi$ decreases, before going through a maximum as the system falls out of equilibrium. The exception is the $\chi_I=0.0$ case, which remains fixed at zero excess helical twist for all densities, consistent with the properties of the {\it Compression} and {\it Defect Crystal} protocols.

\begin{figure}[t!]
\includegraphics[width=3in]{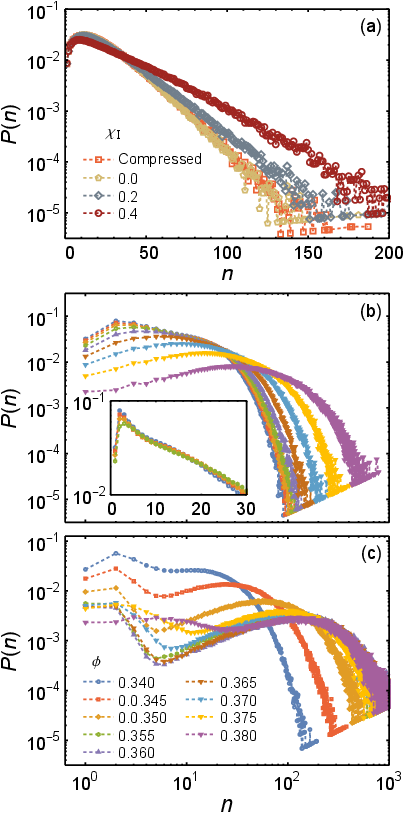}
\caption{\textbf{Excess helix size distribution} (a) Log-linear plot of $P(n)$ as a function of $n$ for the {\it Compressed} simulations and $\chi_I=0.0,0.2,0.4$ at $\phi=0.370$. Log-log plot of $P(n)$ as a function of helix size, $n$, for different $\phi$, for (b) $\chi_I=0.4$ (Insert: Log-Linear plot for the same data at $\phi=0.340-0.355$), (c) $\chi_I=0.8$. Plots (b) and (c) share the same legend for $\phi$. }
\label{fig:excesspn}
\end{figure}

At $\phi=0.370$, the {\it Compression} protocol and {\it Pair} protocols with $\chi_I=0.0,0.2$ and $0.4$ have approximately the same defect fraction, $\langle\theta\rangle\approx 0.085$, but different values of $\langle\chi\rangle$, ranging from zero for the {\it Compression} and $\chi_I=0.0$ cases to about 0.5 for the $\chi_I=0.4$. This suggests the defects must be organised differently, which should be reflected in the size distribution of the helical sections. Figure~\ref{fig:excesspn}(a) shows that $P(n)$ is the same for {\it Compression} and $\chi_I=0.0$, as expected since they both have zero excess helical twist. The distributions for the system with non-zero excess have the same general form, but the maximum moves to smaller $n$, and the larger helical sections become more probable, with increasing $\chi_I$. Figure~\ref{fig:excesspn}(b) shows the density evolution of $P(n)$ for the $\chi_I=0.4$ system. As $\phi$ decreases, the distribution develops a shoulder centred near $n\approx 15$, which is not observed in the zero excess cases. For  $\chi_I=0.8$ (see Fig.~\ref{fig:excesspn}(c)), the distribution develops a deep minimum at small $n$ as $\phi$ decreases, giving rise to a significant peak at $n\approx2$ that indicates the presence of a large fraction of paired defects. Once the system reaches  $\phi=0.355$, where it begins to fall out of equilibrium, the minimum softens as the excess helical twist begins to decrease. The distribution also continues to exhibit a region of power law growth  for small $n$ that crosses over to an exponential decay after going through a maximum at large helical section sizes. The $\chi=0.6$ (not shown) follows the same behaviour, but the minimum in $P(n)$ is not as deep.

\begin{figure}[t]
\includegraphics[width=3in]{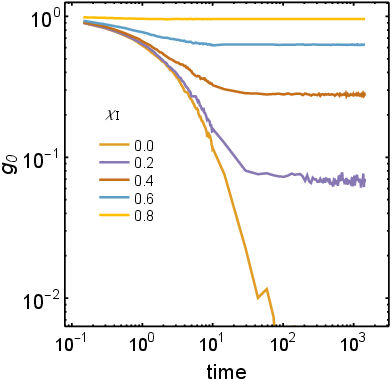}
\caption{\textbf{Excess Orientational Correlation.} Log-log plot of the orientational correlation function, {\it g}$g_0$, as a function of time for different $\chi_I$ at a fixed $\phi=0.360$}
\label{fig:g0excess}
\end{figure}

The presence of an excess in helical twist has a significant effect on the dynamics and structural relaxation of the system. Figure~\ref{fig:g0excess} shows $g_0$ as a function of time for systems with different $\chi_I$, all at $\phi=0.360$.  For the $\chi_I=0.0$ case, $g_0$ decays completely. For $\chi_I>0.0$, $g_0$ follows a similar decay at early times but crosses over to a weak power law decay or plateau at longer times, with a plateau value that increases with increasing excess. The systems with increased excess also have lower pressure, indicating density and free volume are not responsible for the slow dynamics. Structural relaxation in the system is driven by the formation, diffusion and annihilation of defects. However, the decrease in defect fraction associated with increased $\chi_I$ cannot fully explain the slow dynamics. If we compare the relaxation times from the {\it Compression} simulations (Fig.~\ref{fig:oc}(a)), we note that $g_0$ in systems with similar $\langle\theta\rangle$, but with zero excess and higher $\phi$, decay faster than the equivalent systems with excess. This suggests that the defect distribution in the presence of an excess has a significant influence on relaxation. 

\begin{figure*}[p]
\includegraphics[width=6in]{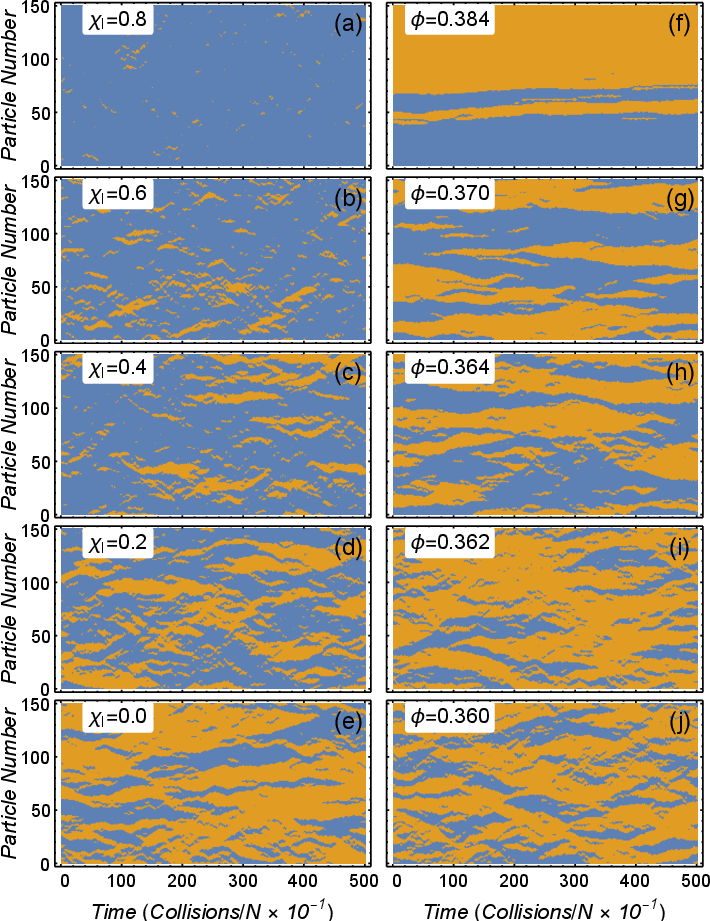}
\caption{\textbf{Helical Twist Trajectory.} Local {\it P} twist (blue) and {\it M} twist (yellow) for particles as a function of time. Left Column (a-e): {\it Pair} trajectories at $\phi=0.360$ for different $\chi_I$. Right Column (f-j): {\it Compression} trajectories at different $\phi$. Neighbouring trajectories in Left and Right columns have similar $\langle \theta \rangle$.}
\label{fig:vistraj}
\end{figure*}

Figure~\ref{fig:vistraj} shows the {\it P} and {\it M} twist directions for individual particles along a series of  trajectories, highlighting the structural and dynamic evolution of the system as a function of the excess in helical twist at a fixed density. We also compare these with the zero excess {\it Compressed} fluid containing a similar defect fraction.  In the presence of a large $\chi_I$ (Fig.~\ref{fig:vistraj}(a)), $\langle\theta\rangle$ is small and defects appear as tightly bound pairs that diffuse together, but have short lifetimes. As a result, structural relaxation involving a change in the local twist direction relies on the creation of nearby defect pairs which give rise to the slow, power law-like decay in $g_0$. The random creation, diffusion and annihilation of the defect pairs throughout the system should disrupt the long-range translational order, suggesting it remains a fluid. As $\chi_I$ decreases, at a fixed $\phi$ (Figs.~\ref{fig:vistraj}(a-e)), $\langle\theta\rangle$ increases and the binding between defects weakens, allowing the formation of larger domains of the minor twist component. Nevertheless, the defects pairs still tend to diffuse together and eventually self-annihilate until $\chi_I=0.2$, where we begin to see longer range diffusion of individual defects. In contrast, Fig.~\ref{fig:vistraj}(b) shows that in the zero excess fluid, at high density and small $\langle\theta\rangle$, defects are isolated so structural relaxation occurs through the diffusion of individual defects. However, we do see evidence of some short lived defect pairs. As $\phi$ decreases and  $\langle\theta\rangle$ increases ((Figs.~\ref{fig:vistraj}(a)), defects become more mobile and appear to diffuse independently. The {\it Compression} and $\chi=0.0$ system becomes equivalent at $\phi=0.360$.

\section{Discussion}
\label{sec:disc}

When confined to a cylindrical, quasi-one-dimensional channel with a diameter $\sqrt{3/4}/7<H_d/\sigma<1+4\sqrt{3}/7$ each particle can only contact the two nearest spheres on either side along the channel and the system forms a simple symmetrical helix at close packing. Despite this simplicity, our simulations show the system exhibits complex thermodynamic, structural and dynamic properties as a function of density that arise from the underlying helical nature of the particle packings and   the inherent importance of defects in the properties of quasi-one-dimensional systems.

The equation of state for the fluid with zero excess helical twist varies continuously and exhibits two shoulders, with corresponding heat capacity maxima, that are related to structural changes in the system. These can be described in terms of how the fluid samples the inherent structure landscape of the system~\cite{Zarif_Inherent_2021}. The low density $C_p$ maximum occurs as the defect fraction plateaus at $\langle\theta\rangle\approx 0.25$, where the average defect separation is four which is the smallest separation required to form a helical section in the inherent structures of the system and is consistent with the  linear to helical crossover observed  in the fluid structure. The high density $C_p$ is located at $\phi\approx 0.375$. This is below the density where the system falls out of equilibrium ($\phi\approx 0.386$), ruling out a glass transition as the possible origin. However, it does coincide with the inflection point of the high density decrease in $\langle\theta\rangle$ (see Fig.~\ref{fig:defects}). The defect fraction provides an approximate measure of the configurational entropy of the fluid which suggests the $C_p$ maximum is related to a Schottky-type anomaly~\cite{Tari_Specific_2003}, where the system rapidly loses configurational entropy in exchange for the increased vibrational entropy of the more ordered inherent structure basins. The inflection point occurs as the system reaches the bottom of the inherent structure landscape where there are few accessible, high density inherent structure basins. A similar effect was identified in the system of quasi-one-dimensional hard discs~\cite{Yamchi_Fragile_2012}.

Some hard particle q1d systems, with degenerate ground states, have been shown to exhibit orientational correlation lengths that diverge at close packing~\cite{Kantor_Universality_2009,Gurin_Orientational_2024}. Figure~\ref{fig:tc} shows that the translational correlation length begins to grow at high density. Unfortunately, our data is not sufficient to determine the nature of the growth in $\xi/\sigma$ because our simulations fall out of equilibrium at $\phi\approx0.386$, which leads to a notable slowdown in the growth of the correlation length. The current system has two degenerate ground states corresponding to the  $\mathsf{M}$ and $\mathsf{P}$ perfect helical packings and it is possible that system spanning fluctuations are needed to flip between the left and right helices as the defect concentration goes to zero and the most dense packed state is approached.
 
The nature of the of the decay of the longitudinal pair correlation function itself has also been of interest. A recent study of q1d hard discs~\cite{Huerta_KT_2020} argues that $|g(z)-1|$ exhibits a power law behaviour at high densities that has its origins in the Kosterlitz-Thouless (KT) transition observed in bulk two-dimensional hard discs. However,  simulations studies that make use of the transfer matrix method to generate equilibrium configurations for large systems show that the power law crosses over to the expected exponential decay at long distances~\cite{ Hu_Comment_KT_2021,Trokhymchuk_ReplyKT_2021}. We see a similar phenomenon here, where the onset of the exponential decay of $|g(z)-1|$ begins at longer distances as $\phi$ increases, which is consistent with the density evolution of $P(n)$, where the small $n$ region described by a power law extends to larger $n$ (Fig.~\ref{fig:hd}), before it becomes exponential.
 
Figure~\ref{fig:oc}(b) shows that the structural relaxation time exhibits a fragile-strong crossover located at the high density $C_p$ maximum. 
Structural relaxation in this system occurs through the creation, diffusion and annihilation of defects that change the local twist direction of the helix. The annihilation of two defects occurs through a spontaneous, irreversible rearrangement of the local structure to produce a singe section of helix. This results in cooperative dynamics and fragile-fluid behaviour at low densities where there are a large number of defects. At high densities, where $\langle\theta\rangle$ is low, structural relaxation occurs through the reversible hopping of isolated defects as they diffuse through the system, which is characteristic of Arrhenius dynamics and strong-fluid behaviour. Similar fragile-strong crossovers, located at the $C_p$ maximum, have a been observed in a variety of systems, including q1d hard discs~\cite{Yamchi_Fragile_2012}, and models of random tetrahedral network forming systems such as silica~\cite{Saika_Free_2004} and ST2 water~\cite{Poole_Dynamical_2011}. A unifying feature of these systems is the presence of well-structured defects that play a central role in structural relaxation, coupled with an inflection in the configurational entropy at high densities or low temperatures. However, the observation of fragile-strong crossovers in molecular glass forming materials remains controversial~\cite{Mallamace_Transport_2010,Chen_On_2012}.

Introducing excess helical twist into the initial starting condition at high density has a dramatic effect on the properties of the system as it is decompressed, with systems given different initial values of $\chi_I$ following distinct equations of state as a function of $\phi$. At high densities the systems are out of equilibrium, generating defects as they slowly relax their almost perfect initial structure. The systems' properties then become time independent at intermediate densities suggesting they have reached some form of local equilibrium. In particular, the systems maintain an excess of helical twist that increases with decreasing density, implying it may be chiral and that the system exhibits a type of topological protection over a range of densities. Eventually the systems fall out of equilibrium at lower density and start to evolve towards the achiral fluid state. Here, it is interesting to note that the large excess states appear to fall out of equilibrium at higher $\phi$, suggesting that if we decompressed the system more slowly it might move to the intermediate excess helical twist states rather than relaxing directly to the achiral fluid. Furthermore, the stability of all the excess helical twist states has disappeared below $\phi\approx 0.33$. At this density $\langle\theta\rangle\approx 0.25$, which corresponds to an average defect separation of four particles, the smallest number of particles required to make a section of a helix.

Jamming phenomena, where particles are trapped in a given structure by geometric constraints, can prevent a system from relaxing to its equilibrium state at high densities. We see an example of this in our {\it Defect Crystal} simulations which, above $\phi=0.386$, exhibit time dependent properties and fail to converge to the same state.
The {\it Pair} simulations, with $\chi=0.0$, provide a control for this effect in our excess helical twist simulations. The initial condition contains a single pair of defects separating two identical sections of helix with opposite twist direction, which gives it the same type of initial condition as the {\it Defect Crystal} simulations with a $\theta=2.0\times 10^{-4}$. The $\chi_I=0.0$ case exhibits the same thermodynamic and dynamic properties as the other zero excess simulations, indicating the system can relax to the equilibrium achiral fluid on the time scale of our simulations in the absence of excess helical twist, which rules out the possibility that jamming is responsible for trapping systems with $\chi_I>0$ in non-equilibrium states.

The topological nature of the defects plays a central role in the stability of the $\chi_I>0$ systems. Introducing an excess helical twist into the initial condition of the system breaks the symmetry of the twist direction observed in the zero excess case. Retaining an excess, distributed through the system, requires defects separating the minor twist component to be paired. Earlier studies~\cite{Zarif_Inherent_2021} of the inherent structures of this system showed that the jamming density, $\phi_J$, of a helical packing containing a single pair of defects increased as the defect separation was decreased. This gives rise to  an entropically driven attraction between defects because increasing $\phi_J$, at a fixed $\phi$ for the fluid, increases the vibrational entropy of the system. Such an attraction could stabilize the formation of the defect pairs observed in the presence of an excess in the helical twist. 

The helix size distribution provides an approximate measure of the radial distribution function for the defects so we can obtain a potential of mean force between defects as $PMF=-\ln P(n)$. The minimum in $P(n)$, observed for large $\chi$ (Fig~\ref{fig:excesspn}(c)), implies there is an attraction between defects at short range and an entropic barrier associated with separating two paired defects, which suggests the topologically protected states may be kinetically stabilized. Defects would remain paired until a fluctuation took them over the barrier allowing them to separate. Over time, as more defects became unbound, the system would evolve towards the achiral fluid state. However, the excess helical twist is time independent at intermediate $\phi$ and $\langle\chi\rangle$ actually increases as the density is decreased. In the $\chi_I=0.8$ case, $\langle\theta\rangle$ also decreases initially with decreasing $\phi$, suggesting the system is not evolving towards the achiral fluid. Finally, at fixed $\phi$, the $PMF$ barrier height decreases as $\chi_I$ decreases, and the minimum in $P(n)$ disappears at $\chi_I=0.4$ (Fig~\ref{fig:excesspn}(b)) indicating the loss of the barrier, but we still see the anomalously slow relaxation in $g_0$. Nevertheless, we cannot rule out the possibility that much longer simulation times may see the topologically protected states eventually decay.

An alternative possibility is that the topologically protected states are stabilized thermodynamically. Hu et al.~\cite{Hu_Correlation_2018} used the transfer matrix method to study the equation of state and correlation lengths of confined hard spheres with next nearest neighbour interactions up to intermediate pressures. They found the largest eigenvalue for the system, which determines the equilibrium properties, consists of two conjugate eigenvalues at low pressure that split into two distinct real eigenvalues at higher pressures. The correlation length associated with the largest eigenvalue continues to grow after the split, while the second largest correlation length goes through a minimum before growing again at higher pressure. Kinks appearing in the correlation lengths, including one associated with the onset of helix formation, do not appear to have signatures in the EOS suggesting the system does not exhibit any phase transitions over the range of densities studied. The EOS of the zero excess fluid in the current study exhibits the same properties as the equilibrium transfer matrix. 

This leaves the possibility that the $\chi>0$ states represent metastable states associated with the second largest eigenvalue that are stabilized through competition between configurational and vibrational entropy. As defects become more tightly bound, the configurational entropy decreases, although the ability to distribute the pairs through the system retains some configurational entropy. However, bringing defects together increases the vibrational entropy. The balance between the number of defects, the degree of defect pairing and the vibrational entropy, along with the density, provides a number variables that could provide different ways of establishing a local equilibrium which may account for the range of states that appear as a function of $\chi_I$ and $\phi$.

The periodic boundaries of the system may also play a role in stabilizing the topologically protected states. Our simulations used translational periodic boundary conditions, but not twist boundary conditions that adjust the helical twist between the periodic images. This imposes a zero twist angle, which may induce a stress on the system at high densities. However, simulations that employed twist boundary conditions~\cite{Fu_Assembly_2017} also found that the perfect helix did not relax to the achiral fluid at intermediate densities.

Finally, bulk hard discs in two-dimensions exhibit a Kosterlitz-Thouless-Halperin-Nelson-Young (KTHNY) transition~\cite{Kosterlitz_Ordering_1973,Halperin_Theory_1978,Young_Melting_1979,Bernard_two_2011,Thorneywork_Two_2017}, which is a continuous transition as a function of density that results from the unbinding of topological defects. Structurally, the unbinding of the defects transforms the high density crystal, which has long-ranged orientational order and quasi-long-ranged translational order to the hexatic phase where the orientational order persists but the translational order has been lost. The KTHNY transition is characterized by the appearance of a power decay in the time dependence of a hexatic orientational order parameter with an exponent that eventually goes to zero in the solid phase. Figure~\ref{fig:g0excess} shows our local twist orientational correlation function, $g_0$ exhibits a similar phenomenology that is tied to the unbinding of  topological defects. Here, $\phi$ is fixed and the order parameter is the excess helical twist. Nevertheless, we see the power law decay appear in the presence of an excess and the slope of the decay decreases as the excess is increased, suggesting this system may exhibit a metastable q1d equivalent to the two dimensional hexatic phase.

\section{Conclusion}
\label{sec:conc}
Our simulations have shown that the thermodynamics, structure and dynamics of a system of confined quasi-one-dimensional hard spheres can in understood in terms of the formation of sections of helical particle arrangements that have alternating twist directions and are separated by topological defects. We show that the zero excess fluid, with equal fractions of spheres with local $\mathsf{P}$ and $\mathsf{M}$, twist directions exhibits two structural crossovers. The first is related to the initial formation of helical structure at low densities. The second occurs as the fluid rapidly eliminates defects in a Schotty-like anomaly, where the system exchanges configurational entropy for the increased vibrational entropy of the more ordered helical states deep in the inherent structure landscape. This high density structural crossover is also associated with a dynamical fragile-strong crossover which is located at the corresponding maximum in the heat capacity. 

We also show that introducing an excess in helical twist into the system leads to the formation of topologically protected states over a range of densities.  These states appear to be stabilized by the interaction between topological defects that become increasingly bound as the amount of excess helical twist is increased. However, the underlying nature of the stabilization still needs to be identified. The potential of mean force, obtained from helix size distribution, reveals the presence of an entropic barrier to defect separation at high excess helical twist that could be responsible for a kinetic stabilization of the defect pairs, but we also find evidence suggesting the topologically protected states could be thermodynamically stabilized. 

The inherent nature of the chirality in these topologically protected states also remains to be studied. For example, our work suggests that colloidal photonic helical structures would be topologically protected, but their chiral properties would be determined by by both the amount of excess helical twist and the way the defects are distributed through the system. Finally, our study has focused on one channel diameter where the ground state is a simple helix. Wider channels form a variety of distinct helical structures with different types of topological defect that may also exhibit topologically protected states.

\section*{Author Contributions}
All authors contributed equally to the conceptualization, writing, reviewing and editing of this manuscript.

\section*{Conflicts of interest}
There are no conflicts to declare.

\section*{Acknowledgements}
We would like to thank the Digital Research Alliance of Canada for computational resources. RKB acknowledges NSERC grant RGPIN--2019--03970 for financial support. MZ  acknowledges supported from the Iran National Science Foundation (INSF) and the Iranian National Foundation of Elites via Grant No. 4015274.


\bibliography{topological}



\end{document}